\documentclass[12pt]{article}\usepackage{graphics}
\newenvironment{no indent}{\noindent\newline---}{}
\begin{document}
\title{New Axial Interactions at a TeV}
\author{\normalsize B. Holdom\thanks{bob.holdom@utoronto.ca}\ \ and T.
Torma\thanks{kakukk@physics.utoronto.ca}\\\small {\em Department of Physics,
University of Toronto}\\\small {\em Toronto, Ontario, M5S1A7,
CANADA}}\date{}\maketitle
\begin{picture}(0,0)(0,0)
\put(310,205){UTPT-98-12}
\put(310,190){hep-ph/9807561}
\end{picture}

\begin{abstract} We consider a heavy fourth family with masses lying in the symmetry
breaking channel of a new strong gauge interaction. This interaction generates a heavy
quark axial-type operator, whose effects can be enhanced through multiple insertions.
In terms of the strength of this operator we can express new negative contributions to
the \(S\) and \(T\) parameters and the shifts of the \(Z\) couplings to the third family. In
particular we find that the new contribution to \(T\) is strongly constrained by the
experimental constraints on the \(Z\) coupling to the \(\tau \).
\end{abstract}
\baselineskip 19pt

\section{Introduction} A heavy fourth family could play a prominent role in dynamical
electroweak symmetry breaking. However, the existence of additional families with
masses greater than a few hundred GeV is thought to be strongly constrained by the
value of the \(S\) and \(T\) electroweak correction parameters. Each new weakly
interacting degenerate fermion doublet contributes \(1/6\pi \) to \(S\). Also at least
some mass splitting in new fermion doublets is difficult to avoid, which for weakly
interacting doublets implies positive contributions to \(T\). The current data, if
anything, favors negative \(S\) and \(T\).

If the fourth family fermions are strongly interacting, it is important to know how these
results for weakly interacting fermions will change. When the new dynamics is
QCD-like it is known that the new fermions will continue to give a positive
contribution to \(S\), since the sign of the corresponding parameter in low energy QCD
is known experimentally \cite{a}. Here we will explore a situation which is distinctly
non-QCD-like in two ways: 1) the strong interaction breaks down close to a TeV, 2) the
fourth family quark masses are in a symmetry breaking channel with respect to this
interaction.

A model based on this type of symmetry breaking pattern has been described in some
detail in \cite{c}. As far as providing some understanding of mass and flavor, the
model has some attractive features in comparison with conventional extended
technicolor models. In particular it was found that the dangerous isospin violating
operators involving the fourth family quarks are naturally suppressed relative to the
operator which feeds mass to the top. (More precisely, the latter operator is enhanced
relative to the former operators.) In addition, the allowed set of effective 4-fermion
operators have a rather different structure than usually considered, and provide a better
chance of producing realistic quark and lepton masses and mixings.   Both of these
features relied on having fourth family quarks in a symmetry breaking channel. It is
therefore worthwhile to understand the implications for precision electroweak
observables, so that experimental data can be brought to bear on this and related
models.

We can comment further on the plausibility of a nonorthodox symmetry breaking
pattern. It is usually assumed that the dynamical mass will occur in the channel
preferred by a single gauge boson exchange (the MAC hypothesis). But it was shown
in \cite{b} that the presence of a gauge boson mass has significant effects on the gap
equation beyond the ladder approximation. As we briefly summarize here, the next
order corrections can be large and of opposite sign to the leading order. Let the strong
interaction be \({\mathit{SU}(N)_{X}}\) (we discuss the cause of its breakdown
shortly). The 4-point kernel appearing in the gap equation for \({N_{f}}\) fermions
transforming as \(N\) or \(\overline{N}\) multiplets, to leading order in \(N\) or
\({N_{f}}\), and after an angular integration, has the form
\begin{equation}\mathit{F}(p, k)=\frac {2}{\pi }(2{C_{2}}(N) - {C_{2}}(R))(\alpha
{F_{1}}(p, k) +  {{\cal{R}}}\alpha ^{2}{F_{2}}(p, k)) + {...}
{\label{a}}\end{equation}
 \(R\) is the \({\mathit{SU}(N)_{X}}\) representation of the condensing channel.
\({{\cal{R}}}\) characterizes the importance of the second order terms, given that
\({F_{2}}\) is appropriately normalized relative to \({F_{1}}\) at the momentum scale
which dominates the integrations. In the case of \({N_{f}}=15\) (which is the case
relevant for the model discussed below) it was found \cite{b} that
\({{\cal{R}}}\approx  - 1.3\) for gauge boson masses corresponding to the breakdown
\({\mathit{SU}(N)_{X}}\rightarrow {\mathit{SU}(N - 1)_{X}}\), with the fermion
mass in the symmetric tensor channel. Given that \(\alpha \) may well be greater than
unity, there is no reason to believe the MAC prediction of a strong repulsion in this
channel.

We view the breakdown of \({\mathit{SU}(N)_{X}}\) as part of the breakdown of a
larger flavor interaction, which also couples to the lighter families. We suppose that
this flavor dynamics produces hierarchies among the various flavor gauge boson
masses, with the lightest being the broken \({\mathit{SU}(N)_{X}}\) gauge bosons.
This hierarchy must be related to the contributions of different order parameters. Since
we restrict ourselves to fermions which carry only standard model quantum numbers,
no fermion bilinear condensates are allowed above the electroweak scale except for
right-handed neutrino condensates.  On the other hand a multitude of 4-fermion
condensates are allowed, and we suppose that some of these are responsible for
\({\mathit{SU}(N)_{X}}\) breaking \cite{c}. For our purposes here we can take the
effective theory describing the \({\mathit{SU}(N)_{X}}\) dynamics to have the gauge
boson masses explicitly present.

We are therefore considering fourth family quarks \(q'\) with dynamical mass in the
\(N\times N\) symmetric tensor representation of a broken \({\mathit{SU}(N)_{X}}\)
gauge interaction. That is, the \({q_{L}^\prime }\) is one component in an
\(\overline{N}\) multiplet and the \({q_{R}^\prime }\) is one component in an \(N\)
multiplet. (We discuss anomaly cancellation below.) The remaining quarks
transforming under \({\mathit{SU}(N - 1)_{X}}\) will not play an important role in
our discussion if they are sufficiently lighter. Given our ignorance of strong
interactions there is also the possibility that \({\mathit{SU}(N)_{X}}\) can be replaced
by a \({\mathit{U}(1)_{X}}\), in which case there is no \({\mathit{SU}(N - 1)_{X}}\)
sector. This possibility will be kept implicit below.

Our approach will be to model the broken gauge interactions as effective 4-fermion
interactions involving \(q'\). This will be reasonable if, in the effective theory, the loops
of interest are dominated by momenta smaller than the 4-fermion compositeness scale.
This requires that there be a well defined separation between compositeness scale and
the fourth family quark masses \({m_{\mathit{q'}}}\ \lower 2pt
\hbox{$\buildrel<\over{\scriptstyle{\sim}}$}\ 1\mathrm{\ TeV}\) (the fourth family
leptons are taken to be somewhat lighter). We might add that our use of 4-fermion
operators to model massive gauge boson dynamics is more justified than their frequent
use in modeling QCD, where the gluons remain massless.

The 4-fermion operators represent the effects of integrating out the massive
\({\mathit{SU}(N)_{X}}\) gauge bosons to all orders in the coupling, and not just the
effects of one gauge boson exchange. These operators must respect the chiral flavor
symmetries of the strong interactions. In fact there is a \(\mathit{SU}(12)\) flavor
symmetry acting on the 12 fields \({q_{L}^\prime }\) and\({(q^{{{\prime
}c}})_{L}}\), accounting for the QCD colors and the up and down flavors. This is not
an exact symmetry; it must be broken by the flavor physics at a higher scale, as well as
by QCD and weak interactions.

To represent the massive gauge boson exchanges, we find that there is only one
independent operator respecting the approximate flavor symmetry. We write it as a
product of color singlet currents,
\begin{equation}\frac {c}{2}(\overline{q}'{\gamma  _{\mu }}{\gamma
_{5}}\mathit{q'})(\overline{q}'\gamma ^{\mu }{\gamma _{5}}\mathit{q'})
{.\label{c5}}\end{equation} We take \(c>0\), which reflects our assumption that the
underlying theory produces an attraction in the \(\overline{q}'q'\) mass channel. This of
course is opposite in sign to the result of the exchange of one massive
\({\mathit{SU}(N)_{X}}\) gauge boson. Our interest in this operator originates in the
fact that one operator insertion in the appropriate 2-loop diagrams will induce
contributions to \(S\) and \(T\) proportional to \( - c\). 

Let \({\hat{\Lambda }}\) be the naive compositeness scale set by the mass of the
\({\mathit{SU}(N)_{X}}\) gauge bosons. We now notice that in the effective theory
below \({\hat{\Lambda }}\) we are able to sum up the effects of the above operator to
leading order in the number of flavors. That is we can sum up multiple insertions of the
operator in the form of bubble chains, illustrated in Fig.\ 1a. Each additional loop is an
axial-vector 2-point function, which at zero momentum will come with a factor
\(4c{f_{\mathit{q'}}^2}\). By \({f_{\mathit{q'}}^2}\) we denote the heavy quark
contribution, which is the dominant contribution, to \(f^{2}\approx (250\mathrm{\
GeV})^{2}\). \({f_{\mathit{q'}}^2}\) includes a factor of \({N_{f}}=6\), the number
of Dirac fermion flavors in the loops. The important point is that these loops are
dominated by momenta below \({\hat{\Lambda }}\), and thus are safely described by
the effective theory.

\begin{center}\includegraphics{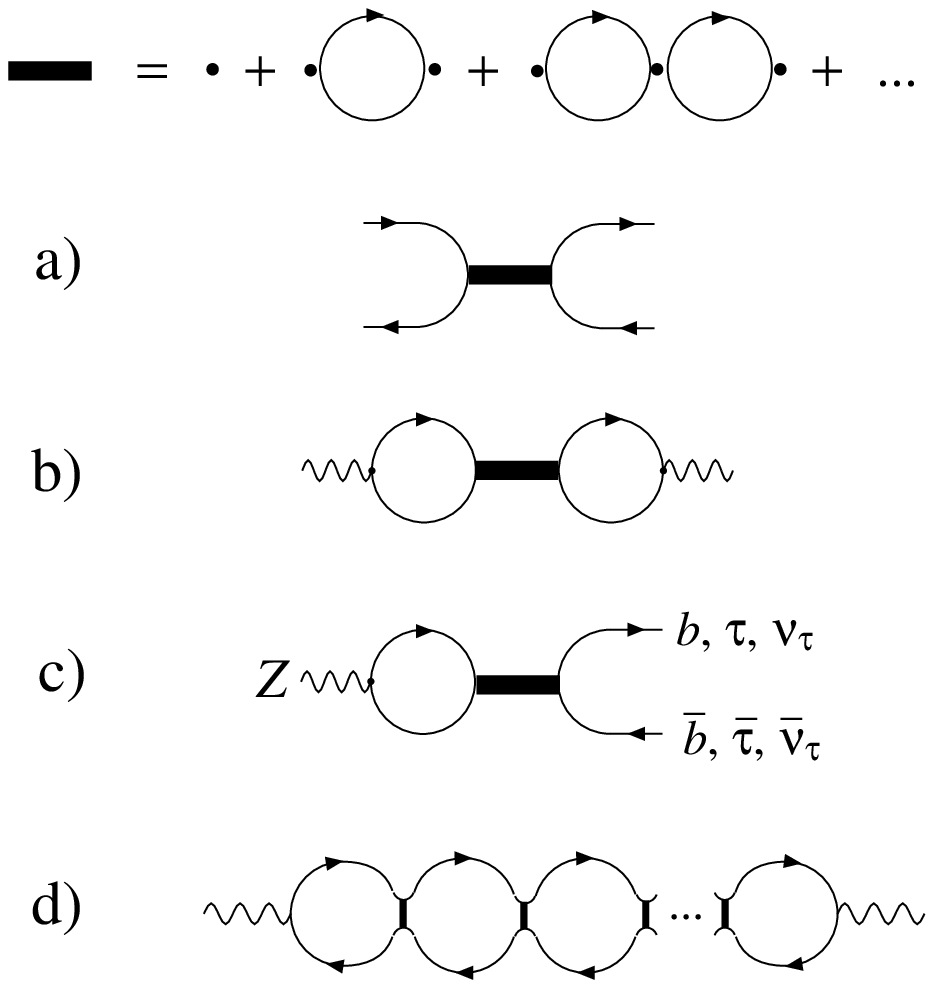}
\end{center} {\small\noindent Figure (1): a) The axial bubble chains modify the
4-fermion operator, at leading order in \(1/{N_{f}}\). b) The axial bubble chains in the
\(T\)-parameter. c) The axial bubble chains in the coupling of the \(Z\) to the third
family. d) The ``chain of chains'' in the \(S\)-parameter.}\\\\One effect of these axial
bubble chains is that the operator in (\ref{c5}) can have an effective compositeness
scale smaller than \({\hat{\Lambda }}\). In other words our operator becomes
effectively nonlocal, of the form
\begin{equation}\overline{q}'{\gamma _{\mu }}{\gamma _{5}}
\mathit{q'}\frac {\mathit{C}(k^{2})}{2}
\overline{q}'\gamma ^{\mu }{\gamma _{5}}\mathit{q'}{\label{ck}}\end{equation}
where \({k_{\nu }}\) is the momentum flowing from one \(\overline{q}'\gamma ^{\mu
}{\gamma _{5}}
\mathit{q'}\) to the other. When \(\mathit{C}(k^{2})\) is calculated and expanded in
\(k^{2}\) we find
\begin{equation}
\mathit{C}(k^{2})={\displaystyle \frac {3}{2}}  {\displaystyle \frac
{{m_{\mathit{q'}}^2}}{{f _{\mathit{q'}}^2}\Lambda ^{2}}} (1 -  {\displaystyle
\frac {k^{2}}{\Lambda ^{2}}}  + {...})\end{equation} where
\begin{equation}\Lambda ^{2}=6{m_{\mathit{q'}}^2}(1{/}(4 c{f_{\mathit{q'}}^2})
- 1).\end{equation}
 \(\Lambda \) becomes the new compositeness scale when \(\Lambda  < {\hat{\Lambda
}}\). In succeeding sections we will let the smaller of  \(\Lambda \) and
\({\hat{\Lambda }}\) be denoted by \(\Lambda \).

In any case, at momentum scales small compared to the true compositeness scale we
have the original local 4-fermion operator with an enhanced coefficient.
\begin{equation}\mathit{C}(0)=\frac {c}{1 - 4c{f_{\mathit{q'}}^2}}\end{equation}
We will restrict ourselves here to \(4c{f_{\mathit{q'}}^2} < 1\), and in particular to the
case when \(\Lambda \) is greater than and not too close to \({m_{\mathit{q'}}}\). We
stress the role played by a large \({N_{f}}\), which can imply a nonnegligible
\(4c{f_{\mathit{q'}}^2}\) even when there is a hierarchy between \({\hat{\Lambda
}}\) and \({m_{\mathit{q'}}}\).

From the form of \(\mathit{C}(k^{2})\) one might wonder whether there is some sort
of tachyonic pole at \(k^{2}\approx  - \Lambda ^{2}\). From the full form of
\(\mathit{C}(k^{2})\) we find that a possible pole is pushed down to more negative
values of \(k^{2}\), in which case it becomes physically meaningless if it occurs in the
region of \( - {\hat{\Lambda }^2}\). There remains the interesting question of what
happens when \(4c{f_{\mathit{q'}}^2}\) is fine-tuned to approach unity from below,
so that a low `mass' tachyonic axial meson apparently does appear. There is also the
possibility of a real axial-vector meson resonance for \(4c{f_{\mathit{q'}}^2}>1\).
These issues will be considered elsewhere\,\cite{bt}.

When Fierz transformed the operator in (\ref{c5}) contains a scalar-scalar
\((\overline{q}'\mathit{q'}) (\overline{q}'\mathit{q'})\) piece with the appropriate sign
to induce a mass. But there is no large \({N_{f}}\) justification (or large \(N\)
justification, since we are discussing only one massive fermion in a
\({\mathit{SU}(N)_{X}}\) multiplet) for the consideration of bubble chains in the
scalar channel, in which case the existence of a light scalar particle with mass of order
the fermion mass becomes questionable. In any case the physics of fermion mass
generation is very sensitive to the physics at scale \({\hat{\Lambda }}\), in contrast to
the physics in the effective theory which we claim is of interest for \(S\) and \(T\). It is
the latter which we focus on in this paper.

\section{The Fourth and Third Families} Our first task is to produce an anomaly free
set of fermions which can realize the previous description. Rather than introducing yet
more fermions, we will instead assume that \({\mathit{SU}(N)_{X}}\) also couples to
the third family in such a way that gauge anomalies are canceled between the third and
fourth families. The consequence of this is another observable effect of the new
physics, in the form of shifts in the \(Z\) couplings to the third family. This in turn will
place strong constraints on how the new physics can affect \(S\) and \(T\).

The third and fourth family quarks transform under
\({\mathit{SU}(3)_{\mathrm{QCD}}}\times {\mathit{SU}(N)_{X}}\) as follows.
\begin{equation}{q_{L}^\prime }=(3, \overline{N}), {\ \ }{q_{R}^\prime }=(3, N), {\
\ }{q_{L}}=(3, N),  {\ \ }{q_{R}}=(3, \overline{N})\end{equation} The broken
\({\mathit{SU}(N)_{X}}\) dynamics is assumed to produce the fourth family quark
masses, corresponding to \({\overline{q}_{1}^\prime }{q_{1}^\prime }\), where the
subscript denotes the first component of an \(N\) or \(\overline{N}\) multiplet. The
third family quark masses corresponding to \({\overline{q}_{1}}{q_{1}}\) are
assumed not to be generated by the broken \({\mathit{SU}(N)_{X}}\) dynamics, even
though their couplings are the same as the fourth family quarks. This is presumably due
to a cross-channel coupling in some effective potential. (The reader may recall how a
two Higgs potential, symmetric between the two Higgs and with a cross-coupling term
\({\phi _{1}^2}{\phi _{2}^2}\), will produce a vacuum expectation value for only one
of the Higgs for a range of parameters.) The \(t\) and \(b\) will instead have masses fed
down via isospin-violating 4-fermion operators generated by flavor physics at a higher
scale. The quarks which transform under \({\mathit{SU}(N - 1)_{X}}\) represent an
additional sector in the theory, and we will return to it below. 

We are in fact discussing a modified version of the model in
\cite{c}, where the \({\mathit{U}(1)_{X}}\) described there is replaced by
\({\mathit{SU}(N)_{X}}\). We refer the reader to that reference for more details on
the generation of quark and lepton masses. A realistic mass spectrum requires that the
charged leptons \(({\tau _{L}^\prime }, {\tau _{R}^\prime } , {\tau _{L}}, {\tau
_{R}})\) transform as \((\overline{N}, \overline{N}, N, N)\) respectively under
\({\mathit{SU}(N)_{X}}\). There are no right-handed neutrinos in the theory at TeV
scales, while the left-handed neutrinos \(({\nu _{L}^\prime }, {\nu _{L}})\) transform
as \((\overline{N}, N)\). Thus the fourth family charged lepton mass \({\overline{\tau
}_{1}^\prime }{\tau _{1}^\prime }\) is in the \(N\times \overline{N}\) channel of
\({\mathit{SU}(N)_{X}}\) while the fourth family Majorana neutrino mass \({\nu
_{L1}^\prime }{\nu _{L1}^\prime }\) is in the symmetric tensor, like the quarks.

We are now able to derive the operators arising from integrating out the massive
\({\mathit{SU}(N)_{X}}\) gauge bosons. We focus on the fourth family (fields with a
prime) and the third family (fields without a prime) and omit the `1' subscripts. There is
now an approximate \(\mathit{SU}(15)\times \mathit{SU}(15)\) flavor symmetry
acting on the 15-plets (\({q_{L}^\prime }\), \({(q^{{{\prime }c}})_{L}}\), \({\tau
_{L}^\prime }\), \({(\tau ^{c})_{L}}\), \({\nu _{L}^\prime }\)) and (\({q_{L}}\),
\({(q^{c})_{L}}\), \({\tau _{L}}\), \({(\tau ^{{{\prime }c}})_{L}}\), \({\nu _{L}}\))
respectively. In addition to this symmetry the effective 4-fermion operators should
respect a discrete symmetry under interchange of these two multiplets. The allowed
operators can then be written in the following form.
\begin{equation}\frac {c^{-}}{2}{J_{5\mu }^-}{J_{5}^{-
\mu }} + \frac {c^{+}}{2}{J_{5\mu }^+}{J_{5}^{+\mu }}{\label{b}}\end{equation}
\begin{eqnarray}&&{J_{\mu }^-}= - \overline{q}'{\gamma _{\mu }}{\gamma
_{5}}\mathit{q'} + \overline{q}{\gamma _{\mu  }}{\gamma _{5}}q + {\overline{\nu
}_{L}^\prime }{\gamma _{\mu }}{\nu _{L}^\prime } - {\overline{\nu }_{L}}
{\gamma _{\mu }}{\nu _{L}} + \overline{\tau }'{\gamma _{\mu }}\tau ' -
\overline{\tau }{\gamma  _{\mu }}\tau \\&&{J_{\mu }^+}=\overline{q}'{\gamma
_{\mu }}{\gamma _{5}}\mathit{q'} + \overline{q}{\gamma _{\mu  }}{\gamma
_{5}}q - {\overline{\nu }_{L}^\prime }{\gamma _{\mu }}{\nu _{L}^\prime } -
{\overline{\nu }_{L}} {\gamma _{\mu }}{\nu _{L}} + \overline{\tau }'{\gamma
_{\mu }}{\gamma _{5}}\tau ' + \overline{\tau }{\gamma _{\mu }}{\gamma
_{5}}\tau \end{eqnarray} There is again a single operator involving the \(\mathit{q'}\)
fields only. \({J_{\mu }^-}\) distinguishes between \(q\) and \(q'\) while \({J_{\mu
}^+}\) does not. Thus it is only the \(c^{-}\) term which causes the \(\overline{q}'q'\)
mass channel to have a different interaction strength than the \(\overline{q}'q\) mass
channel. We expect that \(c^{-} + c^{+}>0\) in order for the \(\overline{q}'q'\) channel
to be attractive, and \(c^{-}>0\) in order for the \(\overline{q}'q'\) channel to be more
attractive than the \(\overline{q}'q\) channel.

There will also be many other operators arising from the flavor physics at a higher
scale, which leave behind the approximate flavor symmetries at a TeV. Some of these
operators may not be negligible at a TeV due to anomalous scaling \cite{c}.  One
example is the operator which feeds mass to the \(t\), and another is the operator
\(({\overline{\tau }_{L}^\prime }{\tau _{R}^\prime })({\overline{\tau }_{R}^\prime
}{\tau _{L}^\prime })\) which can help to induce the \(\tau '\) mass. There are also
some operators (e.g.\ those labeled by \({{\cal{C}}}\) and \({{\cal{D}}}\) in \cite{c})
which contribute to the breakdown of \({\mathit{SU}(N)_{X}}\). But none of these
enhanced operators consist purely of \(q'\) fields. Thus the operators induced by flavor
physics are not expected to compete with the \({\mathit{SU}(N)_{X}}\) induced
operators in our discussion below.

\section{Results} We first consider the contributions to \(T\). We find that the basic
loops can be expressed in terms of the contributions to \(f^{2}\approx (250\mathrm{\
GeV})^{2}\) from the heavy \(q'\) quarks, the \(t\) quark, the \(\tau '\), and the \({\nu
_{L}^\prime }\). We denote these contributions by \({f_{\mathit{q'}}^2}\equiv
{f_{\mathit{t'}}^2} + {f_{\mathit{b'}}^2}\), \({f_{t}^2}\), \({f_{\tau '}^2}\), and
\(2{f_{\nu '}^2}\) respectively. We can represent the various \(f\)'s in terms of effective
ultraviolet cutoffs in the corresponding loops \cite{e}.
\begin{equation}{f_{\mathit{q'}}^2}=\frac {3{m_{\mathit{q'}}^2}}{2\pi
^{2}}\ln(\frac {{\Lambda _{\mathit{q'}}}}{{m_{\mathit{q'}}}}) {,\ \
}{f_{t}^2}=\frac {3{m_{t}^2}}{4
\pi ^{2}}\ln(\frac {{\Lambda _{t}}}{{m_{t}}}) {,\ \ }{f_{\tau '}^2}=\frac {{m _{\tau
'}^2}}{4\pi ^{2}}\ln(
\frac {{\Lambda _{\tau '}}}{{m_{\tau '}}} ){,\ \ }{f_{\nu '}^2}=\frac {{m_{\nu
'}^2}}{4\pi ^{2}}\ln(
\frac {{\Lambda _{\nu '}}}{{m_{\nu '}}}) {\label{fq}}\end{equation} We allow the
various cutoffs to be different, since they are provided either by the compositeness
scale of the 4-fermion operators or by the momentum dependence of the mass in
question.

In terms of these quantities we find
\begin{eqnarray}
\lefteqn{\alpha f^{2}T={\displaystyle \frac {3
\Delta {m_{\mathit{q'}}^2} + \Delta  {m_{{\ell'}}^2}}{ 12\pi ^{2}}}  - {f_{\nu '}^2}
- 4 (c^{-}{\mathit{D}_{1}^2} + c^{+}{\mathit{D}_{2}^2})} \nonumber\\
 & & \mbox{} - 16{f_{\mathit{q'}}^2} (c^{-}{\mathit{D}_{1}} +
c^{+}{\mathit{D}_{2}})({\hat{c}^-}{\mathit{D}_{1}} + {\hat{c}^+}
{\mathit{D}_{2}}){\label{t}}
\mbox{\hspace{1pt}}\end{eqnarray} where
\begin{eqnarray}&&{\hat{c}^\pm }\equiv 
\frac {c^{\pm }}{1 - 4(c^{-} +
c^{+}){f_{\mathit{q'}}^2}}\\&&{\mathit{D}_{1}}=\Delta f^{2} - {f_{t}^2} +
{f_{\nu '}^2}\\&&{\mathit{D}_{2}}=\Delta f^{2} + {f_{t}^2} + {f_{\nu '}^2} -
{f_{\tau '}^2}\\&&\Delta f^{2}={f_{\mathit{t'}}^2} - {f_{\mathit{b'}}^2}\approx
\frac {\Delta {m_{\mathit{q'}}}}{{m_{\mathit{q'}}}}
{f_{\mathit{q'}}^2}\end{eqnarray} and \({m_{\mathit{q'}}}=({m_{\mathit{t'}}} +
{m_{\mathit{b'}}})/2\), \(\Delta {m_{\mathit{q'}}}={m_{\mathit{t'}}}
 - {m_{\mathit{b'}}}\) and \(\Delta {m_{{\ell'}}}={m_{\nu '}} - {m_{\tau '}}\). The
first two terms in (\ref{t}) arise at one loop where the first is the usual contribution and
the \( - {f_{\nu '}^2}\) term arises from the Majorana nature of the \(\nu '\) mass, as
described in \cite{d}. It is clear that \({f_{\nu '}^2}\) must not be too large, if we wish
to avoid fine-tuned cancellations.

The last two terms are the negative contributions from the 4-fermion operators. The
third term is a two loop contribution and the fourth term sums up the axial bubble
chains involving fourth family quarks, as illustrated in Fig.\ 1b. Bubble chains
involving lighter fermions are safely neglected. From (\ref{fq}) we see that these
bubble chains are characterized by a new logarithm for every operator insertion. As we
have mentioned, these chains are leading in \(1/{N_{f}}\), since each bubble sums
over the three colors and two flavors of the
\(q'\) quarks. This chain occurs in the isosinglet channel; isospin breaking masses occur
in the two loops at the ends of the chain, allowing all the interior loops to be without
\({\tau _{3}}\) factors.

Similar corrections are found for the \(Z\) couplings to the third family, as illustrated in
Fig.\ 1c. We find\footnote{The \(Z\) couplings are normalized such that, for example,
\({g_{A}^b}= - 1/2\) and \({g_{L}^{\nu _{\tau }}}=1/2\).}
\begin{eqnarray}&&\Delta {g_{V}^b}=0\\&&\Delta
{g_{A}^b}=2(c^{-}{\mathit{D}_{1}} + c^{+}{\mathit{D}_{2}}) +
8{f_{\mathit{q'}}^2}(c^{-}{\mathit{D}_{1}} + c^{+}{\mathit{D}_{2}})({\hat{c}
^-} + {\hat{c}^+})\\&&\Delta {g_{V}^\tau }=2c^{-}{\mathit{D}_{1}} +
8{f_{\mathit{q'}}^2}(c^{-}{\mathit{D}_{1}} +
c^{+}{\mathit{D}_{2}}){\hat{c}^-}\\&&\Delta {g_{A}^\tau
}=2c^{+}{\mathit{D}_{2}} + 8{f_{\mathit{q'}}^2}(c^{-}{\mathit{D}_{1}} +
c^{+}{\mathit{D}_{2}}){\hat{c}^+}\\&&\Delta {g_{L}^{\nu _{\tau
}}}=c^{-}{\mathit{D}_{1}} + c^{+ }{\mathit{D}_{2}} +
4{f_{\mathit{q'}}^2}(c^{-} {\mathit{D}_{1}} +
c^{+}{\mathit{D}_{2}})({\hat{c}^-} + {\hat{c}^+})\end{eqnarray} These shifts
involve the same two combinations of \({f_{i}^2}\)'s which appeared in the \(T\)
correction. The experimental constraints on the \(Z\) couplings are quite strong
especially for the leptonic couplings, and in particular \(\Delta {g_{A}^\tau }\). This
has the consequence that the 4-fermion contributions to \(T\) are constrained to be
small compared to the first two terms in (\ref{t}).

For \(S\) we find the following result.
\begin{eqnarray}
\lefteqn{S={\displaystyle \frac {15}{24
\pi }}  - {\displaystyle \frac {1}{3\pi
 }} \ln({\displaystyle \frac {{m_{\tau '}}}{{m_{\nu '}}}} )} \nonumber\\
 & & \mbox{} - {\displaystyle \frac {2}{3\pi }} {f_{q'}^2}( {\hat{c}^+} +
{\hat{c}^-})\ln({\displaystyle \frac {\Lambda }{{m_{q'}}}} ) {\displaystyle \frac {1 -
{\displaystyle \frac {1}{6}}  {f_{q'}^2}({\hat{c}^+} + {\hat{c}^-})}{(1 -
{\displaystyle \frac {1}{3}} {f_{q'}^2}({\hat{c}^+} + {\hat{c}^-}))^{2}}}
{\label{s}}\end{eqnarray} The first two terms are the one-loop contributions to \(S\)
from the massive fourth family (with no right-handed neutrino and with all masses
sufficiently above the \(Z\) mass). The origin of the second term is described in
\cite{f}.

The third term is the negative contribution from the 4-fermion operators. Here a loop
integral emerges which is different from the \({f_{q'}^2}\) integral, and this produces a
\(\ln(\Lambda /{m_{q'}})\) dependence in addition to the one in \({f_{q'}^2}\). The
\({\hat{c}^\pm }\)'s appearing in this term again reflect the effect of the axial bubble
chains. The last factor in the third term indicates that we have summed another bubble
chain (in the isovector channel), where at each ``vertex'' the axial bubble chain is 
exchanged in the \(t\) channel. This is illustrated in Fig.\ 1d. This outer chain in the
``chain of chains'' is summing a particular subset of the subleading graphs (subleading
in \(1/{N_{f}}\)). In the case that \(\Lambda  < {\hat{\Lambda }}\), there would be
additional contributions, with loop momenta lying in the range between \(\Lambda \)
and \({\hat{\Lambda }}\). These contributions are complicated by the momentum
dependence of the axial bubble chain (as given by \(\mathit{C}(k^{2})\) in (\ref{ck}));
they could be of the same order as the terms we are keeping although they lack the
\(\ln(\Lambda /{m_{q'}})^{2}\) factor. We have also dropped terms proportional to
any  \({f_{i}^2}\) other than \({f_{q'}^2}\), since the \({f_{q'}^2}\) terms clearly
dominate.

\section{Another Sector?}
\leftmargini=.175in We now consider the additional fermions transforming under the
unbroken, and we assume confining, \({\mathit{SU}(N - 1)_{X}}\). Here the
discussion is much more model dependent. We will just comment on the possible
significant changes to the \(S\) result, since any new contributions to \(T\) are just a
reflection of the unknown up-down mass splittings in the new sector. There are four
possibilities.

\begin{itemize}
\item  The whole \({\mathit{SU}(N - 1)_{X}}\) sector doesn't exist. This would
assume that a \({\mathit{U}(1)_{X}}\) could replace and play the role of the
\({\mathit{SU}(N)_{X}}\) \cite{c}.
\begin{no indent} Our results remain as is.
\end{no indent}
\item  The \({\mathit{SU}(N - 1)_{X}}\) fermion masses are small enough so that the
one loop contributions to \(S\) are small, while the \({\mathit{SU}(N - 1)_{X}}\)
confining scale is large enough so that the bound states have so far escaped detection.
We include here the possibility that the \({\mathit{SU}(N - 1)_{X}}\) fermion masses
are forbidden by certain discrete chiral symmetries\,\cite{g}.
\begin{no indent} There will be additional negative contributions to  \(S\) from
4-fermion operators, for example operators formed as the product of
\({\mathit{SU}(N)_{X}}\) currents, which involve both the \({\mathit{SU}(N -
1)_{X}}\) fermions and the fourth family quarks. The light \({\mathit{SU}(N -
1)_{X}}\) fermions then appear in a loop which only depends logarithmically on the
light fermion mass, giving contributions like the third term in (\ref{s}) except with a
different log factor. These new negative terms can easily be larger than the one
appearing in (\ref{s}).
\end{no indent}
\item  The \({\mathit{SU}(N - 1)_{X}}\) fermion masses are large enough so that the
usual one loop contributions to \(S\) apply, but are still significantly lower than the
fourth family quark masses.
\begin{no indent} We would have to include the positive one loop contributions from 
\({\mathit{SU}(N - 1)_{X}}\) fermions. This could be offset to some extent by terms
like the second term in (\ref{s}), coming from isospin splittings in the quarks and
leptons of that sector. The one loop results are also modified by the \({\mathit{SU}(N -
1)_{X}}\) strong interactions. 4-fermion induced effects beyond those in previous case
would not be substantial as long as the decay constant of the \({\mathit{SU}(N -
1)_{X}}\) fermions is sufficiently less than \({f_{\mathit{q'}}}\).
\end{no indent}
\item  The \({\mathit{SU}(2)_{X}}\) fermion masses are of the same order as the
fourth family quark masses.
\begin{no indent} There would be many additional contributions from the 4-fermion
effects, but the individual fermion masses are reduced because \(f^{2}\) is fixed. The
basic bubble loop contributing to \(S\) would be enhanced relative to those appearing
in \(T\) and the \(Z\) couplings by a \({\mathit{SU}(3)_{X}}\) color factor.
\end{no indent}
\end{itemize}
\section{Discussion} For illustration we provide some numbers for the case when we
ignore all contributions from the model-dependent \({\mathit{SU}(N - 1)_{X}}\)
sector. We note that the strongest constraints come from \(\Delta {g_{V}^\tau
}=.00083\pm .00158\) and \(\Delta {g_{A}^\tau }=.00015\pm .00063\) \cite{h}.

\begin{itemize}
\item  The masses can be fine-tuned,  \({m_{\mathit{q'}}}=670, \Delta
{m_{\mathit{q'}}}=44, {m_{\tau '}}=735, {m_{\nu '}}=304\) GeV with cutoffs 
\({\Lambda _{\mathit{q'}}}={\Lambda _{t}}=2{m_{\mathit{q'}}}, {\Lambda _{\tau
'}}=2 {m_{\tau '}}, {\Lambda _{\nu '}}=2 {m_{\nu '}}\), to give
\(T={\mathit{D}_{1}}={\mathit{D}_{2}}=0\) (and \(f=250\mathrm{\ GeV}\)). Then
\(c^{-}\) and \(c^{+}\) are free to vary to produce a negative \(S\). But this is not a
serious possibility in the absence of a dynamical mechanism to produce these
particular masses.
\item  If we insist on no significant fine tuning among the terms in
\({\mathit{D}_{1}}\) and \({\mathit{D}_{2}}\) then we expect that \( \left|  
{\mathit{D}_{1}}   \right| /f^{2}
\ \lower 2pt \hbox{$\buildrel>\over{\scriptstyle{\sim}}$}\ {0.03}\) and \( \left|  
{\mathit{D}_{2}}   \right| /f^{2}
\ \lower 2pt \hbox{$\buildrel>\over{\scriptstyle{\sim}}$}\ {0.03}\), which in turn
would require that \(c^{-}f^{2}\ \lower 2pt
\hbox{$\buildrel<\over{\scriptstyle{\sim}}$}\ {0.03}\) and \(c^{+}f^{2}\ \lower 2pt
\hbox{$\buildrel<\over{\scriptstyle{\sim}}$}\ {0.02}\). This would imply that the
contributions of the 4-fermion operators to both \(S\) and \(T\) are negligible. The one
loop contributions are such that \(T\) is small as long as there is a suitable hierarchy in
the masses \({m_{\mathit{q'}}}>{m_{\tau '}}>{m_{\nu '}}\) \cite{d}, while
\(S\approx {0.13}({0.08})\) for \({m_{\tau '}}/{m_{\nu '}}=2(3)\).
\item  It has been suggested recently that the \(\tau \)-polarization data shows internal
inconsistencies
\cite{i}. If we exclude the \(\tau \)-polarization data then we have \(\Delta {g_{V}^\tau
}= - .0059\pm .0042\) and \(\Delta {g_{A}^\tau }=.0007\pm .0007\) \cite{h}. For
example the masses \({m_{\mathit{q'}}}\approx 700, \Delta
{m_{\mathit{q'}}}\approx 25, {m_{\tau '}}\approx 550, {m_{\nu '}}\approx 245\)
GeV yield \({\mathit{D}_{1}}/f^{2}\approx {-0.03}\) and
\({\mathit{D}_{2}}/f^{2}\approx {0.03}\) which along with \(c^{-}f^{2}\approx
{0.12}\) and \(c^{+}f^{2}\approx {0.08}\) would produce values of \(\Delta
{g_{V}^\tau }\) and \(\Delta {g_{A}^\tau }\) consistent with the remaining data. In
this case the 4-fermion contribution to \(S\) could easily be comparable to the other
contributions, resulting in a much reduced or even negative \(S\). \(T\) is still
completely dominated by the first two terms in (\ref{t}), which need to cancel at the
10\% level (for these masses and when \(T\approx {-0.2}\)).
\end{itemize} In summary we have considered fourth family quarks in a symmetry
breaking channel of a new strong interaction. Below the symmetry breaking scale we
find a single effective 4-fermion operator involving the fourth family quarks. A single
insertion of this operator generates a negative contribution to both \(S\) and \(T\). We
summed multiple insertions in a complete model in which gauge anomalies are
cancelled through the couplings of the new interaction to the third family. The
constraints on the shifts of \(Z\) couplings to the third family then strongly constrain
the new contributions to \(T\). Significant negative contributions to \(S\) are still
possible.

\section*{Acknowledgments} BH thanks the hospitality and support of the KEK
theory group. TT thanks S. Chivukula for a discussion. This research was supported in
part by the Natural Sciences and Engineering Research Council of Canada.

\end{document}